\documentclass[twocolumn]{aastex63}

\usepackage[]{units}
\usepackage[update,prepend]{epstopdf}
\usepackage{graphicx}
\usepackage{amssymb}
\usepackage{amsmath}
\usepackage{threeparttable}
\usepackage{hyperref}
\usepackage{color}

\usepackage{mathtools}

\usepackage{makecell}

\newcommand{\kms} {\,km\,s$^{-1}$}
\newcommand{\masyr} {\,mas\,yr$^{-1}$}

\newcommand{\Msun}{\,M$_\odot$}

\newcommand{\Fecc}{$F_{\rm ecc}=0.189^{+0.042}_{-0.033}$}
\newcommand{\Feccp}{$F_{\rm ecc}=18.9^{+4.2}_{-3.3}$\%}
\newcommand{\Feccpnumber}{$18.9^{+4.2}_{-3.3}$\%}
\newcommand{\ezero}{$e_0=0.948^{+0.021}_{-0.017}$}
\newcommand{\eone}{$e_1=0.992^{+0.004}_{-0.014}$}

\mathchardef\mhyphen="2D
\shorttitle{Eccentric wide twin binaries}
\shortauthors{Hwang et al.}

\begin{document}

\title{Wide twin binaries are extremely eccentric: evidence of twin binary formation in circumbinary disks}

\correspondingauthor{Hsiang-Chih Hwang}
\email{hchwang@ias.edu}

\author[0000-0003-4250-4437]{Hsiang-Chih Hwang}
\affiliation{School of Natural Sciences, Institute for Advanced Study, Princeton, 1 Einstein Drive, NJ 08540, USA}

\author[0000-0002-6871-1752]{Kareem El-Badry}
\affiliation{Center for Astrophysics $\mid$ Harvard \& Smithsonian, 60 Garden St, Cambridge, MA 02138, USA}

\author[0000-0003-4996-9069]{Hans-Walter Rix}
\affiliation{Max-Planck-Institut für Astronomie, Königstuhl 17, D-69117 Heidelberg, Germany}

\author[0000-0002-5861-5687]{Chris Hamilton}
\affiliation{School of Natural Sciences, Institute for Advanced Study, Princeton, 1 Einstein Drive, NJ 08540, USA}

\author[0000-0001-5082-9536]{Yuan-Sen Ting}
\affiliation{Research School of Astronomy \& Astrophysics, Australian National University, Cotter Rd., Weston, ACT 2611, Australia}
\affiliation{School of Computing, Australian National University, Acton ACT 2601, Australia}

\author[0000-0001-6100-6869]{Nadia L. Zakamska}
\affiliation{School of Natural Sciences, Institute for Advanced Study, Princeton, 1 Einstein Drive, NJ 08540, USA}
\affiliation{Department of Physics \& Astronomy, Johns Hopkins University, Baltimore, MD 21218, USA}

\begin{abstract}

The {\it Gaia} mission recently revealed an excess population of equal-mass ``twin'' wide binaries, with mass ratio $q\gtrsim 0.95$, extending to separations of at least 1000\,AU. The origin of this population is an enigma: twin binaries are thought to form via correlated accretion in circumbinary disks, but the typical observed protostellar disks have radii of $\sim100$\,AU, far smaller than the separations of the widest twins. Here, we infer the eccentricity distribution of wide twins from the distribution of their $v$-$r$ angles, i.e., the angle between the components' separation and relative velocity vectors. We find that wide twins must be on extremely eccentric orbits. For the excess-twin population at 400-1000\,AU, we infer a near-delta function excess of high-eccentricity system, with eccentricity $0.95 \lesssim e \leq 1$. These high eccentricities for wide twins imply pericenter distances of order $10$\,AU and suggest that their orbits were scattered via dynamical interactions in their birth environments, consistent with a scenario in which twins are born in circumbinary disks and subsequently widened. These results further establish twin wide binaries as a distinct population and imply that wide twins can be used as a probe of the dynamical history of stellar populations. 

\end{abstract}

\section{Introduction}

Binary population demographics encode information about the star formation process and subsequent dynamical processing. One striking feature of the binary population is the existence of an excess population of equal-mass ``twin'' binaries with mass ratios $0.95 \lesssim q \leq 1$, where $q=M_2/M_1$ and $M_1$ and $M_2$ are the masses of the primary and secondary. This twin excess is strongest at short periods, and was long thought to exist only at orbital periods $P_{\rm orb} \lesssim 40$ days \citep[e.g.][]{Lucy1979twin, Tokovinin2000}. However, studies of large samples of binaries have recently shown that an excess population exists even at very wide separations, extending beyond 1000 AU for solar-type stars \citep[e.g.][]{ Soderhjelm2007, Moe2017,El-Badry2019}.

The leading hypothesis for the origin of the twin excess is that twins form in circumbinary disks. In this scenario, the lower-mass star has a higher accretion rate from the circumbinary disk due to its wider orbit around the common center of mass, or due to the dynamics of the accretion streams \citep{Bate1997,Farris2014, Young2015b,Duffell2020}, thus driving the mass ratio to unity. Torques from the disk may also shrink the binary orbit, potentially explaining why the amplitude of the twin excess is largest at close separations \citep{Tokovinin2020}. However, this picture is still under debate. First, it is unclear whether the lower-mass star in a circumbinary disk will always have the higher accretion rate, with different codes making qualitatively different predictions \citep{Bate1997, Ochi2005} and the behavior may depend on gas temperature \citep{Young2015a, Young2015b}. Furthermore, how the interaction between the binary and the circumbinary disk affects a binary's orbital evolution remains an open question \citep{Artymowicz1991,Artymowicz1994,Pichardo2005, Shi2012,Miranda2017,Munoz2019, Moody2019, Ragusa2020,Tiede2020,Heath2020, Dittmann2022}. Thus, twin binaries provide a unique observational opportunity to study the effects of accretion from circumbinary disks.

The excess of twin binaries with separations $\gtrsim 100$\,AU is particularly puzzling \citep{Soderhjelm2007}. Their existence out to $>1000$\,AU is now confirmed by {\it Gaia} and is not due to selection effects \citep{El-Badry2019}. Observed protoplanetary and circumbinary disks have typical radii of $\sim$100\,AU \citep[e.g.][]{Andrews2018disk,Ansdell2018,Manara2019}, and so wide twins cannot have formed in circumbinary disks at their current separations.
Although the excess twin population only contains a few percent of all binaries at these separations, the twin excess is manifest as a sharp, step function-like jump in the mass ratio distribution above $q_{\rm twin}\approx 0.95$ \citep{El-Badry2019}. Because $q_{\rm twin}$ does not vary between close and wide binaries, it is natural to assume that wide twins form via the same process as close twins. One possible formation scenario is that wide twins formed at closer separations ($10-100$\,AU) within a circumbinary disk and were subsequently widened by dynamical interactions. However, the nature of this interaction and widening process is poorly understood.

If twins' formation involves dynamical widening either through gravitational interactions with other stars, or through recoils or kicks to one of the components, then highly eccentric orbits are expected for wide twins. Furthermore, binary eccentricities are a direct prediction from simulations of binary-circumbinary disk interaction \citep[e.g.][]{Artymowicz1994,Roedig2011,Munoz2019,Dittmann2022}. Therefore, the eccentricity of wide twin binaries is a critical connection between observational and theoretical work.

In this letter, we investigate the eccentricities of wide twin binaries. The individual eccentricities of wide binaries are challenging to measure due to their long orbital periods ($\gtrsim10^3$\,yr for binaries at $>10^2$\,AU), but the population eccentricity distribution can be statistically constrained by the distribution of $v$-$r$ angles, the angle between the separation vector ($r$) and the relative velocity vector ($v$) of a wide binary \citep{Hwang2022ecc}. This paper is structured as follows. Sec.~\ref{sec:sample} explains the sample selection. Sec.~\ref{sec:result} presents the main results, showing that wide twin binaries are highly eccentric. We discuss the results and conclude in Sec.~\ref{sec:conclusion}.

\section{Sample selection}
\label{sec:sample}

\begin{figure}
	\centering
	\includegraphics[width=1\linewidth]{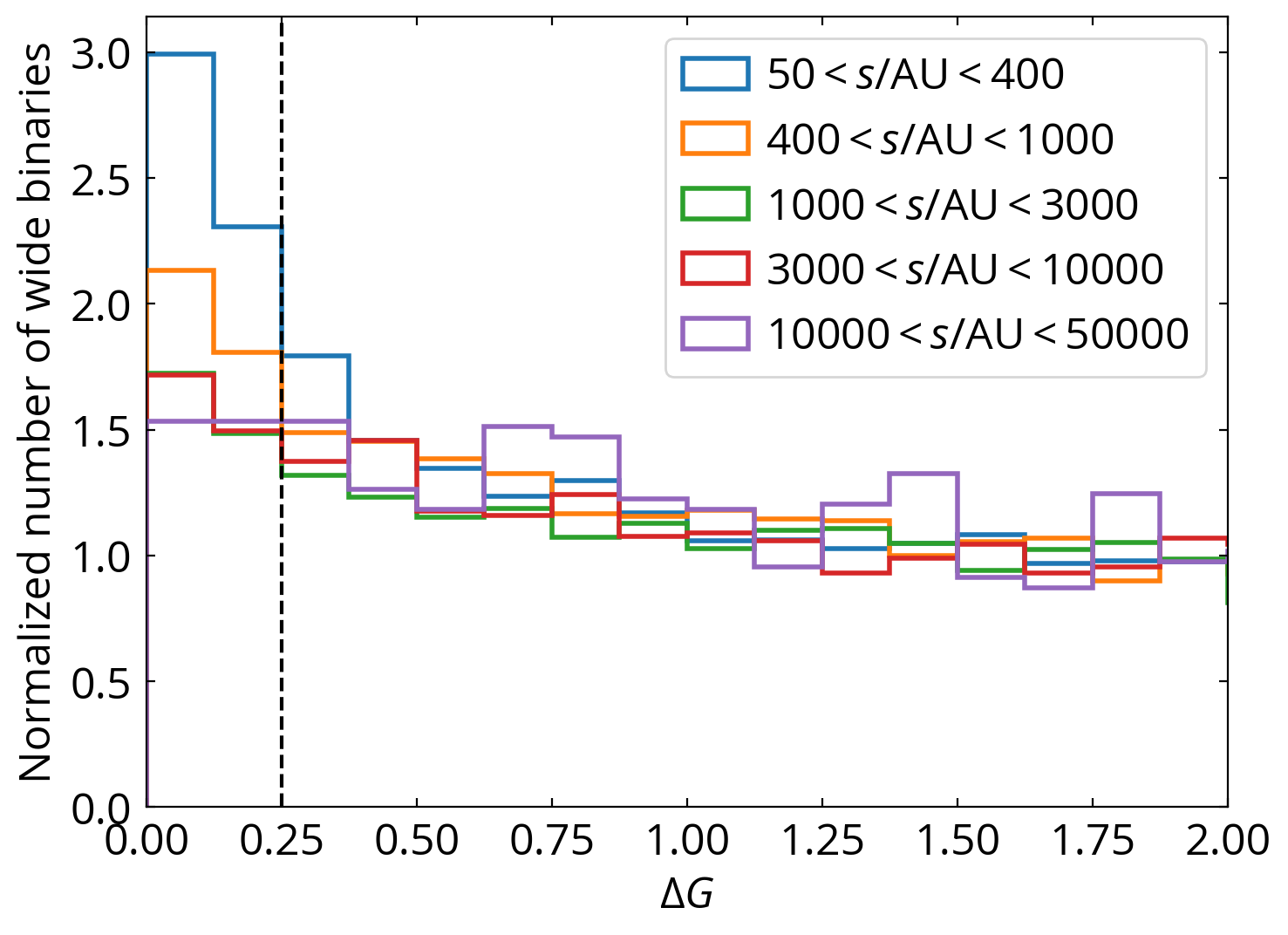}
	\caption{The distributions of magnitude difference $\Delta G$ among the components of wide binaries at different separations. The vertical dashed line marks $\Delta G=0.25$\,mag, the selection adopted for \emph{twin} wide binaries in this paper. The distributions are normalized to unity at $\Delta G=1.5$-2\,mag. Compared to binaries at $>10000$\,AU, binaries at separations $\lesssim3000$\,AU show a significant excess of twins at $\Delta G<0.25$\,mag.  }
	\label{fig:deltaG}
\end{figure}

The high-precision astrometry from {\it Gaia} \citep{Gaia2016} has enabled large-scale wide binary search \citep{Oh2017, El-Badry2018b, Tian2020, Hartman2020, Hwang2021a}. In this work, we use the $\sim1$-million wide binaries within 1\,kpc \citep{El-Badry2021} selected from {\it Gaia} early Data Release 3 \cite[eDR3,][]{Gaia2021Brown}. To avoid contamination, we require the chance-alignment probability $<0.1$ \citep{El-Badry2021}.

We use the $v$-$r$ angle method to infer the eccentricity distribution of wide binaries \citep{Tokovinin1998,Tokovinin2016, Tokovinin2020a,Hwang2022ecc}. Qualitatively, the $v$-$r$ angle distribution of randomly-oriented circular binaries peaks at 90$^\circ$, and the peak moves toward 0$^\circ$ and 180$^\circ$ for more eccentric orbits (e.g. Fig. 3 in \citealt{Hwang2022ecc}). Quantitatively, \cite{Hwang2022ecc} developed a Bayesian framework to derive the posterior of the underlying eccentricity distribution given a set of observed $v$-$r$ angles, i.e. $p(\{\alpha_j\}|\{\gamma_{i}\})$, where $\{\alpha_j\}$ are the free parameters of the eccentricity distribution and $\{\gamma_i\}$ is a set of $v$-$r$ angles. We refer the reader to \cite{Hwang2022ecc} for the detail of the Bayesian framework.

The observed $v$-$r$ angles of {\it Gaia} wide binaries are computed from the angle between projected $v$ and projected $r$ vectors, where $r$ is the vector connecting two member stars' sky coordinates and $v$ is the vector of two stars' proper motion differences. We require that all wide binaries in the sample have angular separations $>1.5$\arcsec\ to avoid {\it Gaia}'s systematics on $v$-$r$ angles in pairs below $1.25$\arcsec\ \citep{Hwang2022ecc}. To reliably measure $v$-$r$ angles, we further limit the sample to have parallaxes $>5$\,mas (i.e. distances $<$200\,pc) and proper motion differences that are $3\sigma$ from being zero. With {\it Gaia}'s proper motion precision of $\sim0.1$\masyr, simulations in \cite{Hwang2022ecc} show that these selection criteria can recover 89.7\% of 1000 AU wide binaries' proper motion differences at $>3\sigma$ at distances of 200\,pc, assuming that they are equal-solar-mass binaries with the so-called thermal eccentricity distribution ($f(e)de=2ede$). All $v$-$r$ angle measurements and related quantities used in this work are publicly available from \cite{Hwang2022ecc}.

With these selection criteria, the classification based on the absolute $G$-band magnitudes and $BP$-$RP$ colors (\texttt{binary\_type} in \citealt{El-Badry2021}) shows that 88\%\ of the wide binaries with separations of 3000-10000\,AU (hence angular separations $>15$\arcsec) are double main-sequence (MS) binaries, 10\%\ are white dwarf (WD)-MS, and the rest are double-WD wide binaries. In {\it Gaia} eDR3, $G$-band photometry has an angular resolution of $\sim0.7$\arcsec\,\citep{Gaia2021Fabricius}. However, because $BP$ and $RP$ photometry does not have deblending treatment, $BP$ and $RP$ are not reliable for pairs $<2$\arcsec. For this reason, we do not use $BP$-$RP$ colors to differentiate MS from WD in this work, but we expect the sample to be dominated ($\gtrsim 90$\%) by MS-MS wide binaries. 

Following \cite{El-Badry2019}, we select wide twin binaries by $\Delta G < 0.25$\,mag, where $\Delta G$ is the difference in {\it Gaia}'s broadband photometry $G$ of two component stars of a wide binary. For MS-MS binaries, $\Delta G < 0.25$\,mag corresponds to a mass-ratio selection of $q > 0.95$, with some slight dependence on the primary mass \citep{El-Badry2019}. Fig.~\ref{fig:deltaG} shows the distributions of $\Delta G$ for different binary separations. Compared to binaries at $>10000$\,AU, wide binaries $\lesssim3000$\,AU have an excess of twin binaries at $\Delta G<0.25$\,mag. The median error of $\Delta G$ in our sample is 0.001\,mag, much smaller than the width of the excess twin component at $\Delta G \sim 0.25$\,mag.

Binaries selected by $\Delta G<0.25$\,mag contain two components: one component is from the extension of the smooth, power-law-like mass-ratio distribution at $q\gtrsim0.5$, and the other component is from the {\it excess} twins that only contribute to $q>q_{twin}\sim0.95$ \citep{Moe2017,El-Badry2019}. Here we use binaries at separations $>10000$\,AU as the baseline where no significant excess twins are present \citep{El-Badry2019} to quantify the fraction of excess twins at other separations. Specifically, binaries with $\Delta G < 0.25$\,mag constitute $p_{400-1000}=7.70\pm0.16$\% of all wide binaries at 400-1000\,AU, compared to $p_{>10000}=5.89\pm0.36$\% at $>10000$\,AU. Therefore, at $\Delta G<0.25$\,mag and binary separations of 400-1000\,AU, the fraction of {\it excess} twins is $\mathcal{E}_{400-1000}= (p_{400-1000} - p_{>10000})/p_{400-1000} = 23.5\pm 5.1$\%. Similarly, $\mathcal{E}_{1000-3000}=4.0\pm4.8$\% for separations at 1000-3000\,AU, where the excess is more significant for a certain primary mass range \citep{El-Badry2019}. Due to the difference in the detailed sample selection (e.g. using $BP$ and $RP$ photometry or not, different distance distribution, and lack of corrections for incompleteness), the derived twin excess $\mathcal{E}$ may differ from the intrinsic mass-ratio distribution models inferred by \cite{El-Badry2019}.

\begin{figure}
	\centering
	\includegraphics[width=1\linewidth]{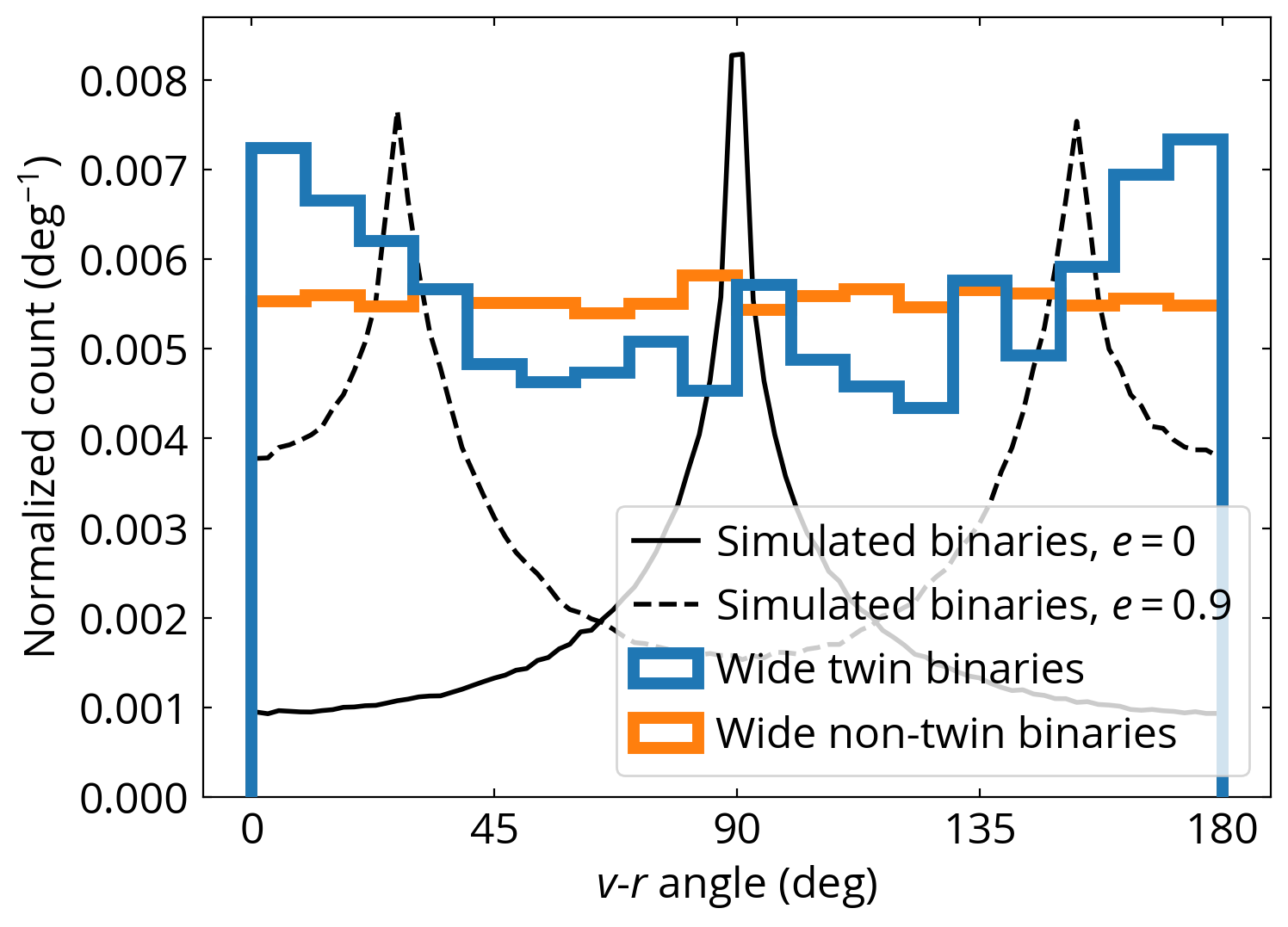}
	\caption{The $v$-$r$ angle distributions for wide binaries with separations at 400-1000\,AU. The solid and dashed black lines are simulated binaries with arbitrary normalization, showing that circular orbits (solid) have a $v$-$r$ angle distribution peaking at $90^\circ$, and the peaks for eccentric $e=0.9$ orbits (dashed) are closer to $0^\circ$ and $180^\circ$. The $v$-$r$ angle distributions of wide twin binaries (blue) are strongly enhanced at 0$^\circ$ and 180$^\circ$ compared to non-twins (orange), indicating the presence of highly eccentric twins with $e>0.9$.}
	\label{fig:vrangle}
\end{figure}

\begin{figure}
	\centering
	\includegraphics[width=1\linewidth]{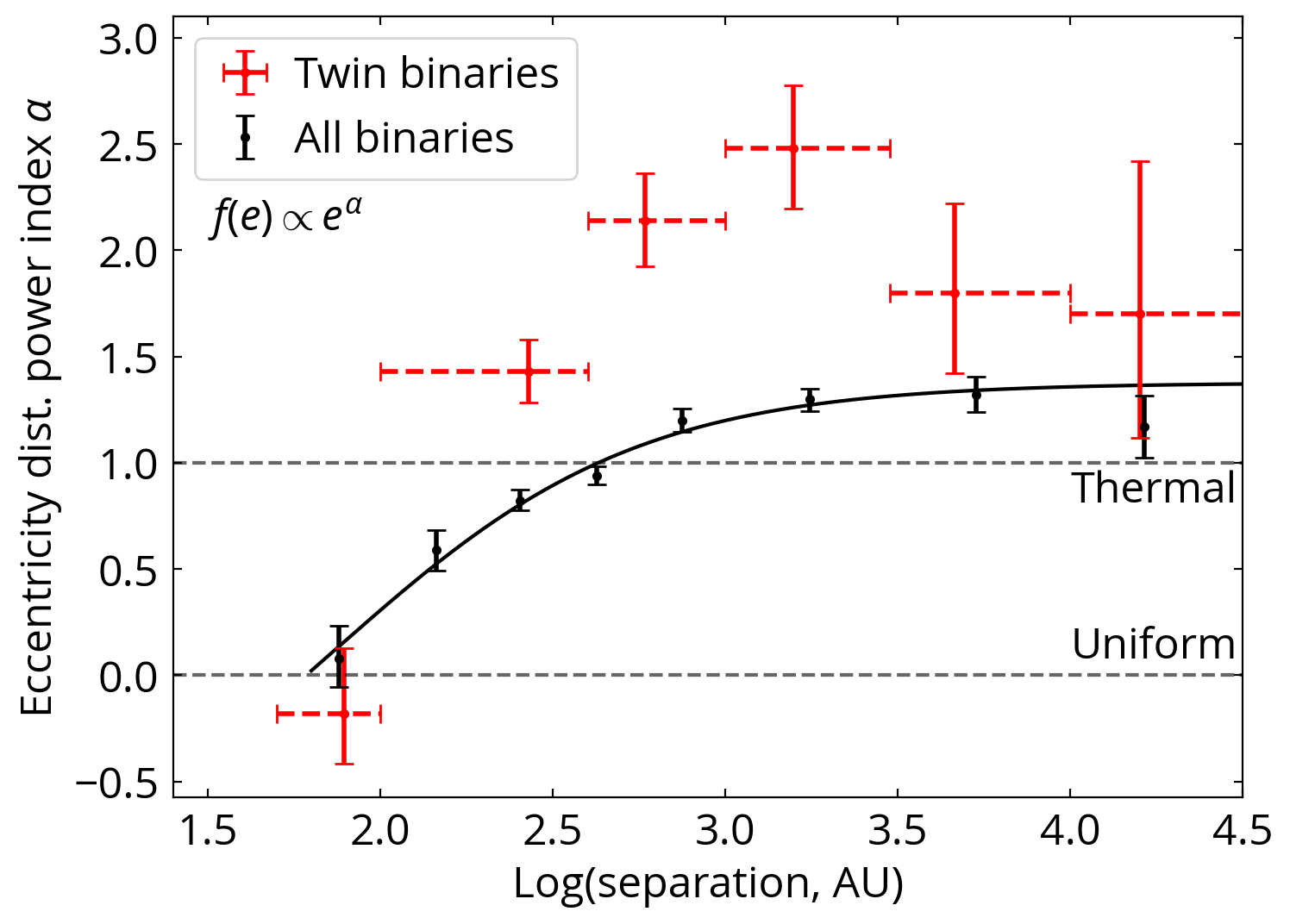}
	\caption{The power-law indices $\alpha$ of eccentricity distributions ($f(e)\propto e^{\alpha}$) as a function of binary separations. The black symbols are the results from \cite{Hwang2022ecc} for all binaries, which are dominated by non-twins; the red markers show $\alpha$ and separation bin sizes (horizontal) for the twin binaries. At 100-3000\,AU, twin binaries are significantly more eccentric (larger $\alpha$) than other wide binaries at the same separations. } 
	\label{fig:alpha-sep}
\end{figure}

\begin{figure*}
	\centering
	\includegraphics[height=0.35\linewidth]{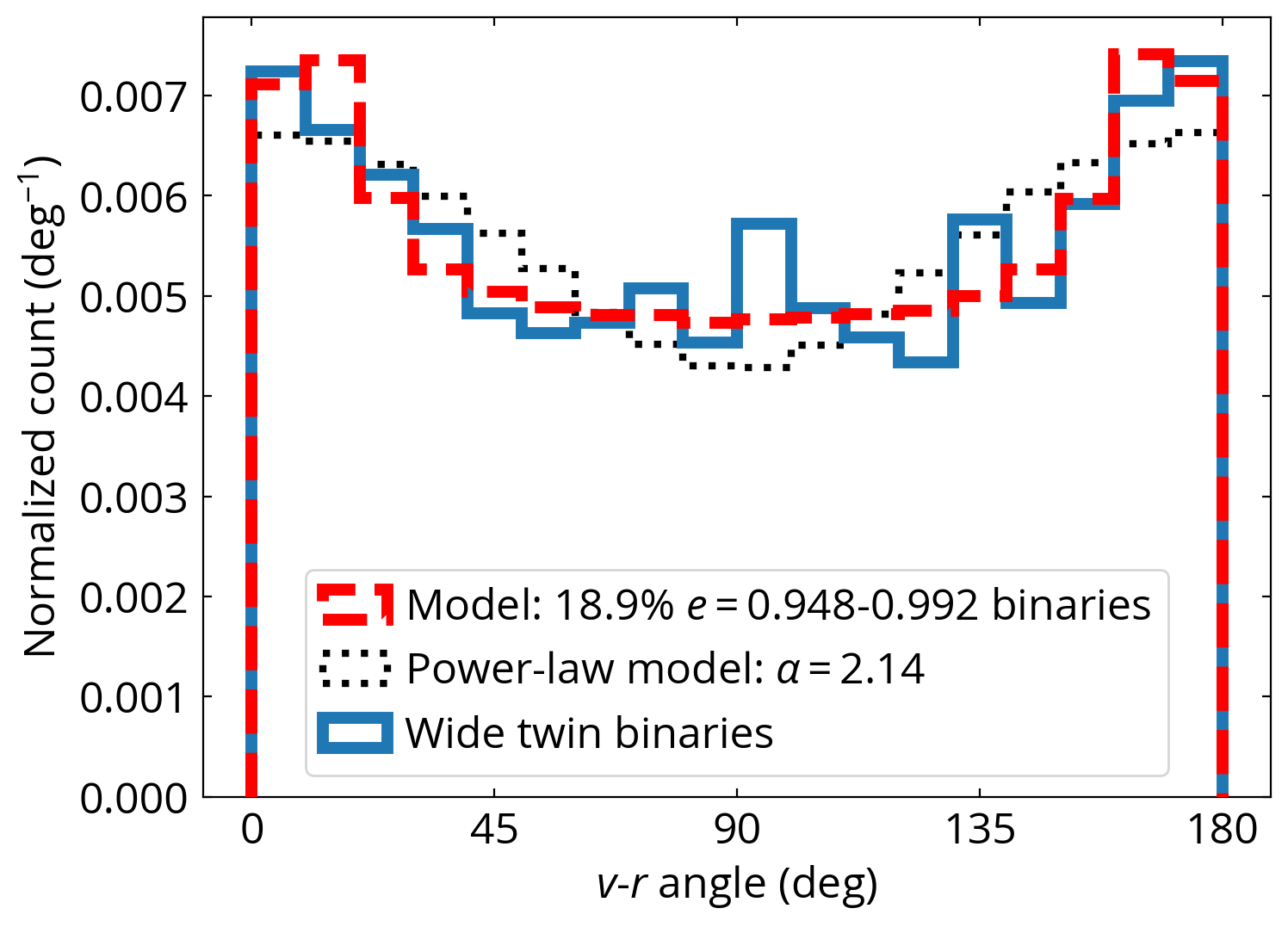}
	\includegraphics[height=0.35\linewidth]{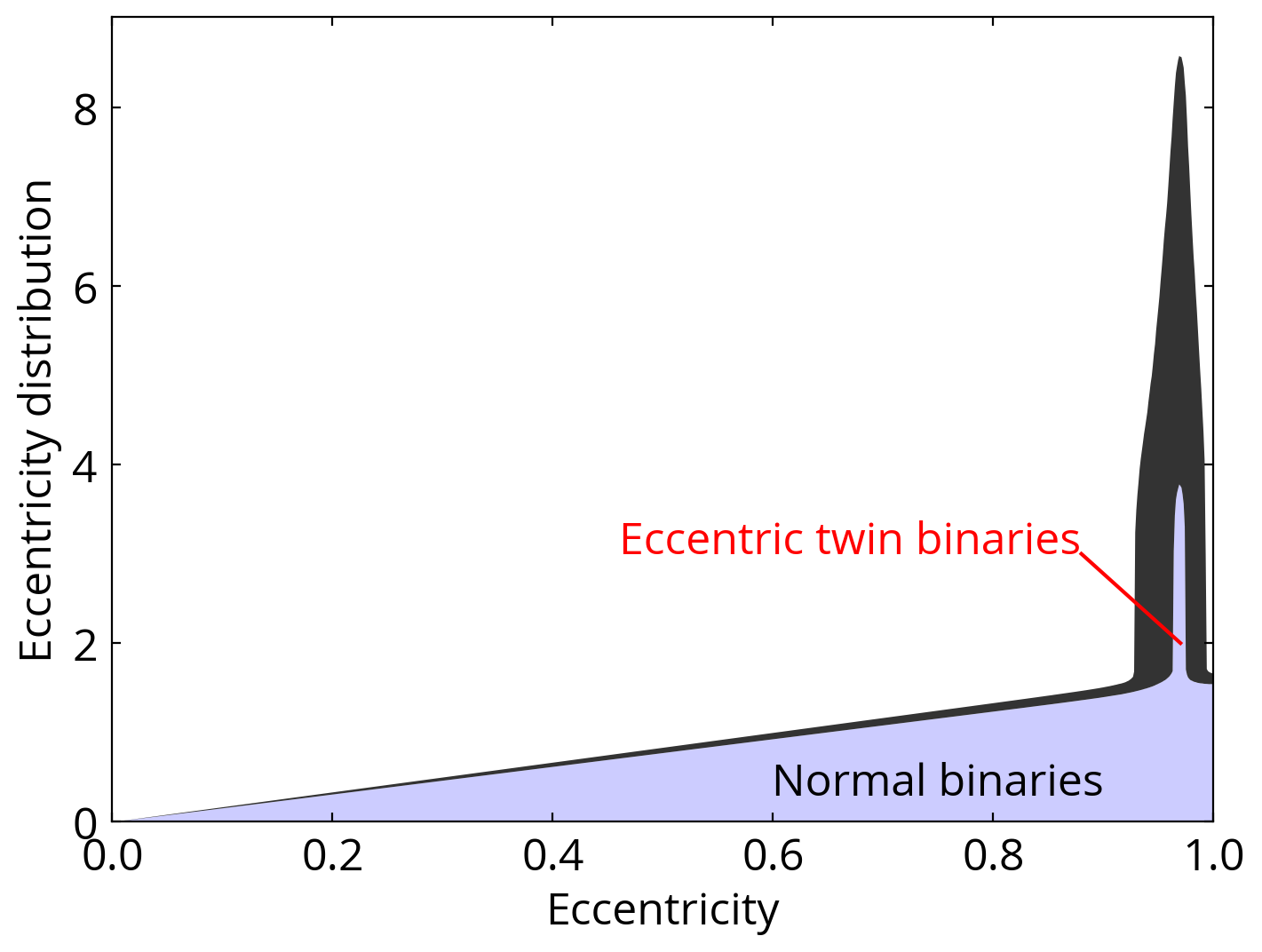}
	\caption{Left: the $v$-$r$ angle distributions for wide binaries at 400-1000\,AU (blue) and for simulated models (black and red). The best-fit power-law model of $\alpha=2.14$ (dotted black) well describes the overall trend of twin binaries (solid blue) but with some noticeable differences. The dashed red line corresponds to 81.1\% of normal wide binaries and 18.9\% of $e=0.948$-$0.992$ binaries, which agrees with the observation better than the power-law model. Right: the modeled eccentricity distribution (the dashed red line in the left panel) for twin binaries at separations 400-1000\,AU. The black shaded region shows the 20-80 percentiles of the model. } 
	\label{fig:model}
\end{figure*}

\section{Wide twin binaries are eccentric}
\label{sec:result}

Fig.~\ref{fig:vrangle} shows the distribution of $v$-$r$ angles for wide binaries with binary separations of 400-1000\,AU. Wide twin binaries (blue) are selected by $\Delta G<0.25$\,mag, and wide non-twin binaries (orange) by $\Delta G > 0.25$\,mag. Fig.~\ref{fig:vrangle} also shows that the simulated $v$-$r$ angle distributions for circular binaries (solid black) with random orientation strongly peaks at $90^\circ$, and the peaks move toward $0^\circ$ and $180^\circ$ for more eccentric orbits (e.g. $e=0.9$ for the dashed black line). The $v$-$r$ angles of wide twin binaries are strongly enhanced at 0$^\circ$ and 180$^\circ$, indicating the presence of highly eccentric binaries with $e>0.9$. In contrast, wide non-twin binaries in this separation range have a flat $v$-$r$ angle distribution, suggesting an underlying thermal eccentricity distribution \citep{Hwang2022ecc}.

Using the Bayesian inference detailed in \cite{Hwang2022ecc}, we measure the eccentricity distributions of wide twin binaries as a function of projected binary separations (denoted by $s$). Specifically, we model the eccentricity distribution $f(e)$ as a power law $f(e)=(1+\alpha)e^\alpha$, and obtain the best fit $\alpha$ given the observed $v$-$r$ angle distribution.

Fig.~\ref{fig:alpha-sep} shows the best-fit power-law indices $\alpha$ for the eccentricity distributions of twin binaries (red) as a function of binary separations. For comparison, the black points and solid black line show the results from \cite{Hwang2022ecc} for all wide binaries, which are dominated by non-twin MS-MS binaries. If one excludes twin binaries in the black points from \cite{Hwang2022ecc}, then their $\alpha$ values are only modified by an amount smaller than their measurement uncertainties. The red dashed horizontal error bars indicate the bin sizes, and the vertical error bars represent the 68\%\ credible interval. At separations of 100-3000\,AU, twin binaries have significantly larger $\alpha$ than wide non-twin binaries with similar separations, meaning that they are highly eccentric. There is no significant difference in $\alpha$ between twins and non-twins at $>3000$\,AU, in agreement with the fact that there is no significant twin excess at larger separations \citep{El-Badry2019}. Interestingly, at 50-100 AU, twin binaries do not have significantly different $\alpha$ even though the excess twin population is very significant at these small separations \citep{El-Badry2019}, hinting that the majority of twin binaries below 100\,AU do not experience the orbit-widening processes that alter their orbital eccentricities and they may have formed in disks at their current separations. We remind the reader that ``twin binaries'' in Fig.~\ref{fig:alpha-sep} includes both ``excess'' twins and the smooth background population, implying that the excess twin are even more eccentric than suggested by Fig.~\ref{fig:alpha-sep}.

The left panel in Fig.~\ref{fig:model} shows the $v$-$r$ angle distribution of wide twin binaries at 400-1000\,AU and the simulated binaries with the best-fit $\alpha=2.14$ (dotted black line). Although the overall observed distribution agrees with the best-fit power-law result, there are some subtle but significant differences. Specifically, compared to the observed distribution (blue), the power-law model (black) is lower at 90$^\circ$, 0$^\circ$, and 180$^\circ$, and is higher at 45$^\circ$ and 135$^\circ$. These differences suggest that the eccentricity distribution of wide twin binaries is not a perfect power law.

Alternatively, we can model the eccentricity distribution of wide twin binaries as a sum of two populations: ``normal'' binaries that just happen to have $q\approx 1$, and excess twins. We assume that a fraction $F_{\rm ecc}$ of twins have unusually eccentric orbits and the remaining $1-F_{\rm ecc}$ of them follow the eccentricity distribution of non-twin binaries at similar separations. Therefore, the total eccentricity distribution $f(e)$ of wide twin binaries is

\begin{equation}
\label{eq:f-e}
    f(e) = (1-F_{\rm ecc})f_{\rm normal}(e) + F_{\rm ecc} f_{\rm ecc}(e),
\end{equation}
where $f_{\rm normal}(e)$ and $f_{\rm ecc}(e)$ are the eccentricity distributions for normal binaries and the excess eccentric binaries, respectively. At 400-1000\,AU, the non-twin ($\Delta G >0.25$\,mag) wide binaries have a flat $v$-$r$ angle distribution and its best-fit $\alpha$ is $1.05\pm0.05$, and therefore we adopt a thermal eccentricity distribution ($\alpha=1$) for $f_{\rm normal}(e)=2e$ \citep{Hwang2022ecc}. For excess eccentric binaries, we choose $f_{\rm ecc}(e)$ as a top-hat function whose value is $1/(e_1-e_0)$ between two free parameters $e_0$ and $e_1$, and zero elsewhere. With details shown in Appendix~\ref{sec:mcmc}, we use the affine invariant Markov chain Monte Carlo (MCMC) ensemble sampler \texttt{emcee} \citep{Foreman-Mackey2013} to constrain $F_{\rm ecc}$, $e_0$, and $e_1$.

The best-fit parameters are \Fecc, \ezero, \eone. The best-fit values are the most probable values of the marginalized posterior distributions, and the uncertainties represent the highest posterior density interval that includes 68 per cent of the area. Therefore, the observed $v$-$r$ angle distribution is best fit by \Feccp\ of highly eccentric ($e>e_0=0.948$) binaries among the twin binaries. The red dashed histogram in Fig.~\ref{fig:model} shows the simulated $v$-$r$ angle distribution of this model, well consistent with the observed distribution. The best fit of $f(e)$ (Eq.~\ref{eq:f-e}) in the right panel of Fig.~\ref{fig:model} illustrates the presence of the highly eccentric twin binaries.

The median error of the $v$-$r$ angles for 400-1000\,AU twin binaries is 3.3$^\circ$, meaning that the highest measurable eccentricity is $e=\cos(3.3^{\circ})=0.9983$ \citep{Hwang2022ecc}. Therefore, our measured \eone\ is approaching the precision limit. Despite the high eccentricities close to the hyperbolic regime ($e>1$), the symmetric observed $v$-$r$ angle distribution in Fig.~\ref{fig:model} indicates that they are still on stable Keplerian orbits; otherwise, disrupting binaries on hyperbolic orbits would have $v$-$r$ angles enhanced at $0^\circ$ but not at 180$^\circ$.

One natural explanation is that all excess twins identified from the mass-ratio distribution are highly eccentric. In this case, we would expect the twin excess fraction from the mass-ratio distribution equals the fraction of the eccentric component, i.e. $\mathcal{E}=F_{\rm ecc}$. At separations 400-1000\,AU, the fraction of eccentric twins \Feccp\ is consistent with the excess twin fraction at $\Delta G<0.25$\,mag measured from the $\Delta G$ (i.e. mass ratio) distribution, $\mathcal{E}_{400-1000}=23.5\pm 5.1$\% (Sec.~\ref{sec:sample}). However, with the similar analysis applying to twin binaries at 1000-3000\,AU, the fraction of eccentric twins is $F_{ecc,1000-3000}=21.6^{+5.9}_{-4.5}$\%, which is much higher than $\mathcal{E}_{1000-3000}=4.0\pm 4.8$\%. Therefore, the connection between the excess in the mass ratio distribution ($\mathcal{E}$) and the excess in the eccentricity distribution ($F_{\rm ecc}$) remains not fully established, and future work is needed to investigate if $F_{\rm ecc}$ follows the same primary mass dependence as $\mathcal{E}$ \citep{El-Badry2019}.

\begin{figure}
	\centering
	\includegraphics[width=1\linewidth]{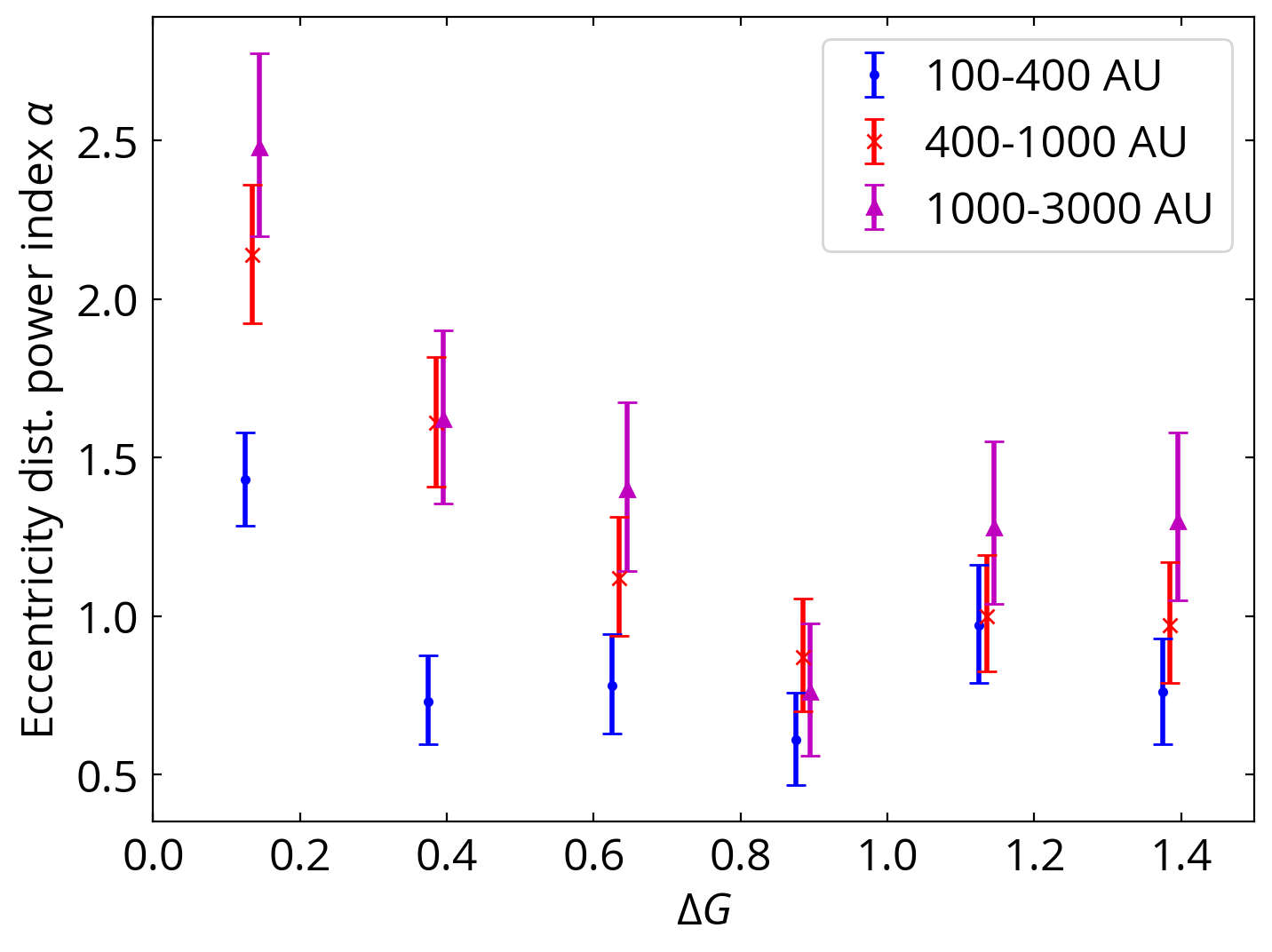}
	\caption{The power-law indices $\alpha$ of eccentricity distributions ($f(e)\propto e^{\alpha}$) as a function of $\Delta G$. Only $\Delta G<0.25$\,mag has a significant increase in $\alpha$, suggesting that the high eccentricity is specifically associated with the twin excess, instead of a general trend with mass ratios. } 
	\label{fig:alpha-dG}
\end{figure}

Fig.~\ref{fig:alpha-dG} shows the eccentricity distribution power-law indices $\alpha$ versus $\Delta G$. For all separation bins between 100 and 3000\,AU, $\alpha$ strongly increases at $\Delta G<0.25$\,mag, the criterion used to select twin binaries. 
The marginal increase in $\alpha$ at $\Delta G=0.25-0.5$ for binaries at 400-1000\,AU may hint that this bin still has some contribution from the eccentric twin binaries. At $\Delta G>0.5$\,mag, $\alpha$ becomes flat for all separation bins. Therefore, the high-eccentricity population is specifically for twin binaries ($\Delta G< 0.25$\,mag), and the eccentricity distribution does not strongly depend on mass ratios anymore at $\Delta G>0.5$\,mag ($q\lesssim0.9$).

\section{Conclusions and Discussion of formation channels}
\label{sec:conclusion}
Wide twin binaries are a mysterious population that has near equal-mass ratios ($q>0.95$), common among close binaries, but has large binary separations of $\sim1000$\,AU. In this letter, using the $v$-$r$ angle method to infer the eccentricity distribution \citep{Hwang2022ecc}, we show that these wide twin binaries are highly eccentric (Fig.~\ref{fig:vrangle}, \ref{fig:alpha-sep}). Our result suggests that \Feccpnumber\ of wide twin binaries (selected by $\Delta G<0.25$\,mag) at 400-1000\,AU have eccentricities between \ezero\ and \eone (Fig.~\ref{fig:model}). Despite the high eccentricities, their symmetric $v$-$r$ angle distribution with respect to $90^\circ$ suggests that they are on stable Keplerian orbits, instead of being dissolving binaries. This high eccentricity is specific to excess twins, as opposed to being a smooth function of the mass ratio (Fig.~\ref{fig:alpha-dG}).

Binaries on eccentric orbits will most likely be observed near their apocenters where they spend most of their time due to the lower orbital velocities. For apocenters $r_{apo}=1000$\,AU, the inferred eccentricities of $e=0.95$ and $0.99$ correspond to pericenters of $r_{peri}=r_{apo}(1-e)/(1+e)=26$\,AU and 5\,AU, respectively. The high eccentricities of $\lesssim1000$\,AU wide twins must originate from their formation, since interactions with passing stars and secular torquing by the Galactic tide mainly affect wide binaries at $\gtrsim10^4$\,AU, and the latter have been shown only make the initial eccentricity distribution closer to thermal \citep{Hamilton2022}. Pericenters smaller than the typical $\sim100$\,AU sizes of circumbinary disks are consistent with the scenario in which twins are born with initial separations smaller than the size of the circumbinary disk and they subsequently widen through the dynamical interactions in the birth environments.

The nature of the dynamical interactions causing the wide and eccentric orbits remains uncertain. One approach to excite high eccentricities is an instantaneous velocity kick. To change a circular orbit with an orbital velocity $v_{circ}$ to an eccentric orbit with $e>0.9$, the required velocity kick $v_{kick}$ is comparable with the original circular velocity ($v_{kick}=v_{circ} (1-\sqrt{1-e})$), where $v_{circ}=19$\kms\ at 5\,AU. Therefore, an instantaneous kick with $v_{kick}\sim10$\kms\ taking place in close binaries can produce eccentric ($e>0.9$) wide binaries. However, the source of such strong velocity kicks is currently not known in star formation processes. Protostellar outflows and jets can reach velocities of several $100$\kms, but they are launched from the disk at the stage when the disk mass is only $\sim10$\%\ of the central star \citep{Bally2016}, and thus are unlikely to impose a strong kick on the star.

Chaotic three-body interactions can lead to the formation of close binaries with highly eccentric tertiary companions \citep{Reipurth2012}. If the close binary was unresolved, the resulting triple system would look like a wide binary. However, radial velocity variations among wide twin binaries as well as the flux excess due to the unresolved companions suggest that unresolved companions are not more common in wide twins than in wide non-twins \citep{El-Badry2019}. Therefore, wide twin binaries are unlikely to form from chaotic three-body interaction.

For hierarchical triples that are not formed from chaotic three-body interaction, the resolved outermost companions would have less eccentric orbits required by dynamical stability \citep{Shatsky2001, Tokovinin2016,Hwang2022ecc}. Since the presence of unresolved companions can cause non-zero $\Delta G$, the equal-mass selection by $\Delta G<0.25$\,mag preferentially excludes systems with unresolved companions, thus allowing more eccentric outer companions. In other words, we are investigating the possibility that wide twins are more eccentric than wide non-twins because unresolved companions are more common in wide non-twins. This scenario is possible because it is not uncommon to have unresolved companions in (predominantly non-twin) wide pairs \citep{El-Badry2018, Hwang2020c, Fezenko2022}. To test this effect, we simulate the photometry of wide binaries using \texttt{MIST} \citep{Dotter2016,Choi2016} and \texttt{brutus}\footnote{\url{https://github.com/joshspeagle/brutus}} (Speagle et al. in prep), where 50\% of them are assumed to have unresolved companions and all component stars' masses are drawn from the Kroupa initial mass function \citep{Kroupa2001IMF}. We find that $\Delta G>0.25$\,mag is only $\sim5$\% more likely to have unresolved companions than $\Delta G<0.25$\,mag (for reference, $\Delta G>0.25$\,mag corresponds to a 1-\Msun\ primary with an $>0.79$-\Msun\ unresolved secondary. Note that the effect on $\Delta G$ from an unresolved companion is different from the previous case of resolved binaries). Therefore, this potential effect from the lack of unresolved companions in wide twins is not able to explain \Feccp\ of high-eccentricity binaries at $\Delta G<0.25$\,mag.

Twin binaries may form through the enhanced accretion and interaction with circumbinary disks \citep{Tokovinin2020}. Then the strong interaction between binaries and the circumbinary disk may increase binary's eccentricity, even though the observed $e>0.95$ is unusually high for typical disk-binary interaction in simulations \citep{Artymowicz1991, Cuadra2009, Roedig2011}. Alternatively, the process that widens binary orbits and causes high eccentricities may not be specific to twins. It is possible that such a process takes place among all close binaries, producing both eccentric wide twins and non-twins out to $\sim1000$\,AU. Then since twins are more common in close binaries, the high-eccentricity component is more apparent in wide twins than wide non-twins. Future investigations are needed to establish the connection between close binaries, wide binaries, and their mass ratios and eccentricities.

Our results suggest that wide twin binaries have eccentricities approaching unity. Therefore, the process making these twin binaries wide and eccentric should also disrupt some of binaries during the star formation. These disrupted binaries may contribute to the low-mass runaway or walkaway stars in star-forming regions (e.g. \citealt{Schoettler2020}), predicting a population of equal-mass runaway or walkaway pairs with opposite directions. Furthermore, some of these wide twin binaries may have pericenters comparable to or smaller than the radii of giant stars, which may lead to collision at later stellar evolution and the formation of blue stragglers \citep{Kaib2014}, although the twin fraction is lower in more massive stars where they can evolve to giants within the Hubble time \citep{Moe2017, El-Badry2019}.

\section*{Acknowledgements}

The authors are grateful to the referee for the constructive report. HCH appreciates the discussions with Jim Stone, Roman Rafikov, and Scott Tremaine. HCH acknowledges the support of the Infosys Membership at the Institute for Advanced Study. HWR acknowledges support from the GIF grant I-95-303.5-2018. This work was supported by a grant from the Simons Foundation (816048, CH). Y.S.T. acknowledges financial support from the Australian Research Council through DECRA Fellowship DE220101520. NLZ is supported at the IAS by the J. Robert Oppenheimer Visiting Professorship and the Bershadsky Fund.

\appendix

\restartappendixnumbering
\setcounter{figure}{0}

\section{Markov-Chain Monte-Carlo results}
\label{sec:mcmc}

Our eccentricity model has three free parameters: $F_{\rm ecc}$, $e_0$, and $e_1$ in Eq.~\ref{eq:f-e}. According to Bayes' theorem, their posterior distributions are 
\begin{equation}
\label{eq:bayes}
    p(\{F_{\rm ecc}, e_0, e_1\}|\{\gamma_{obs,i}\}) \propto \\ p(F_{\rm ecc}) p(e_0) p(e_1) \Pi_i \int p(\gamma_{obs,i} | \gamma_{true,i}) p(\gamma_{true,i}| e_i)  p(e_i|\{F_{\rm ecc}, e_0, e_1\}) d\gamma_{true,i} de_i,
\end{equation}
where $i$ is the index of individual binaries, $\gamma_{obs,i}$ is the observed $v$-$r$ angle for binary $i$, $p(\gamma_{obs,i} | \gamma_{true,i})$ is the error distribution for $v$-$r$ angles. To probe the high eccentricity values close to 1, we use binary simulations to compute the $p(\gamma_{true,i}|e_i)$ for eccentricities from 0 to 1 with a step of 0.001 and $\gamma_{true,i}$ with a step of $0.5^{\circ}$. The details of these terms can be found in \cite{Hwang2022ecc}.

\begin{figure}
	\centering
	\includegraphics[width=0.4\linewidth]{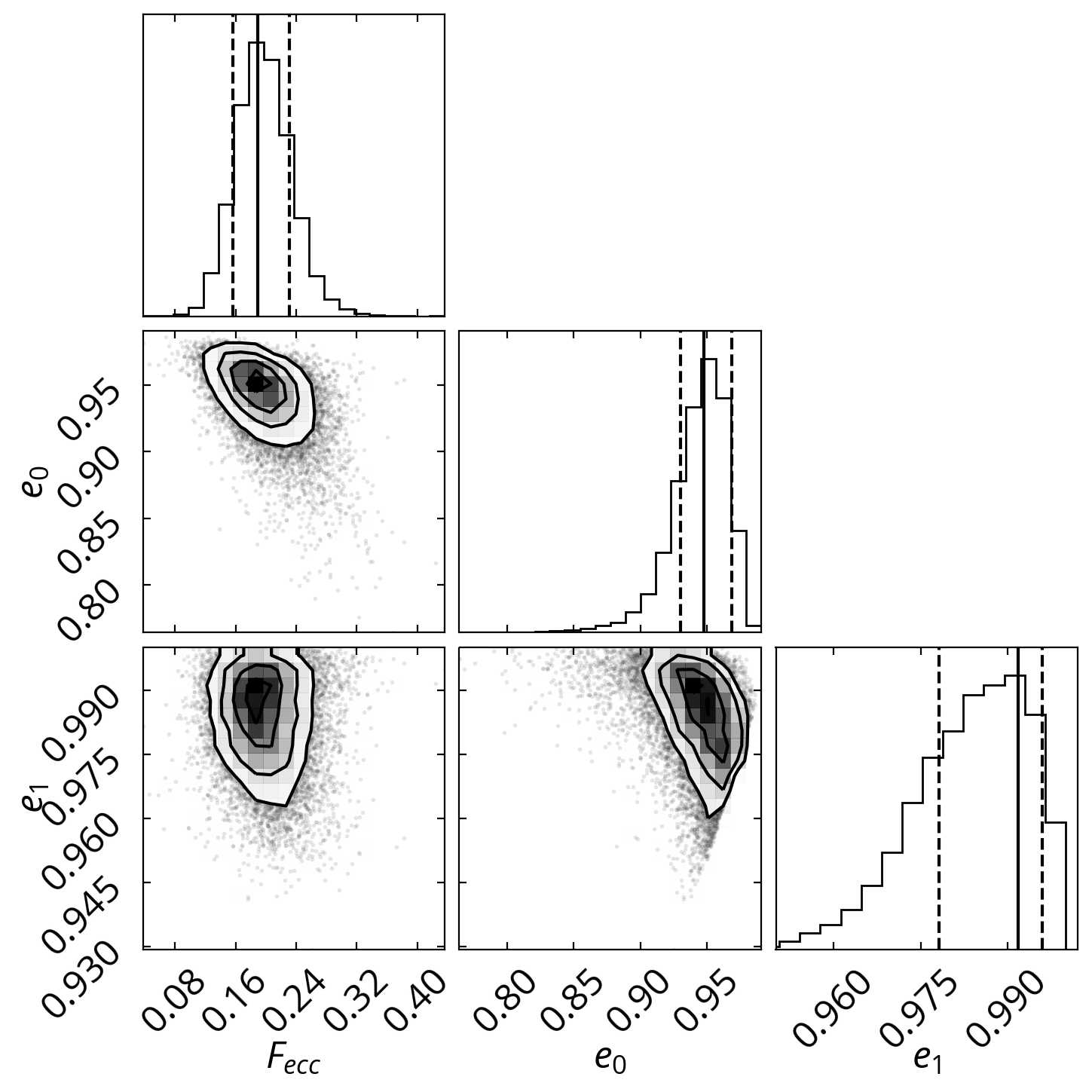}
	\caption{The posterior distributions of the eccentricity distribution model. The solid lines are the most probable values of the marginalized distributions, and the dashed lines are the highest posterior density interval that includes 
	64-per cent of the area.}
	\label{fig:corner}
\end{figure}

We then use \texttt{emcee} to derive the posterior distributions. In the MCMC run, we numerically compute the two-dimensional integral in Eq.~\ref{eq:bayes} with equal spacings of $\Delta e=0.001$ and $\Delta \gamma=0.5^{\circ}$. Uninformative flat priors are adopted for $p(e_0)$, $p(e_1)$, and $p(F_{\rm ecc})$, and we further require that $e_1>e_0$. We use the Gaussian move as the proposal function in the MCMC to mitigate the effect of finite $\Delta e$ in the integral calculation. The resulting posterior distributions for 400-1000\,AU wide twin binaries are shown in Fig.~\ref{fig:corner} \citep{Foreman-Mackey2016}.

\bibliography{paper-TwinBinaryEcc}{}
\bibliographystyle{aasjournal}
\end{document}